\documentclass[conference]{IEEEtran}
\IEEEoverridecommandlockouts
\usepackage{cite}
\usepackage{amsmath,amssymb,amsfonts}
\usepackage{algorithmic}
\usepackage{graphicx}
\usepackage{textcomp}
\usepackage{xcolor}
\usepackage{multirow}
\usepackage{array}
\usepackage{url}
 \usepackage{booktabs}
\def\BibTeX{{\rm B\kern-.05em{\sc i\kern-.025em b}\kern-.08em
    T\kern-.1667em\lower.7ex\hbox{E}\kern-.125emX}}
\begin{document}

\title{Fusing Spectral Correlation Density Imaging with Deep Learning for Intelligent Fault Diagnosis in Rotating Machinery \\
}

\author{\IEEEauthorblockN{1\textsuperscript{st} Given Name Surname}
\IEEEauthorblockA{\textit{dept. name of organization (of Aff.)} \\
\textit{name of organization (of Aff.)}\\
City, Country \\
email address or ORCID}}

\author{\IEEEauthorblockN{Dilshara Herath\IEEEauthorrefmark{1}, Chinthaka Abeyrathne\IEEEauthorrefmark{1}, Chamindu Adithya\IEEEauthorrefmark{2}, Chathura Seneviratne\IEEEauthorrefmark{1}}
\IEEEauthorblockA{\textit{\IEEEauthorrefmark{1}Department of Electrical and Information Engineering, \IEEEauthorrefmark{2}Department of Mechanical and Manufacturing Engineering} \\
\textit{\IEEEauthorrefmark{1}\IEEEauthorrefmark{2}Faculty of Engineering, University of Ruhuna, Galle, Sri Lanka} 
}

\IEEEauthorrefmark{1}dilshara.herath3@gmail.com,
    \IEEEauthorrefmark{2}dilhanchinthaka99@gmail.com,
    \IEEEauthorrefmark{3}csadithya1@gmail.com,
    \IEEEauthorrefmark{4}chatura@eie.ruh.ac.lk
}

\maketitle

\begin{abstract}

Bearing fault diagnosis in rotating machinery is critical for ensuring operational reliability, therefore early fault detection is essential to avoid catastrophic failures and expensive emergency repairs. Traditional methods like Fast Fourier Transform (FFT) often fail to capture the complex, non-stationary nature of vibration signals. This study leverages the cyclostationary properties of vibration data through Spectral Correlation Density (SCD) images to enhance fault detection and apply deep learning for classification. Using a publicly available dataset with bearing faults seeded in two distinct housings (A and B) under varying load conditions (0 Nm, 2 Nm, 4 Nm), we processed vibration signals into 2D SCD images to reveal fault-specific periodicities, such as broadband spectra (2000–8000 Hz) for larger faults. Three convolutional neural network (CNN) models, Custom CNN, ResNet152V2, and EfficientNetB0, were developed to classify seven bearing conditions. The custom CNN achieved the highest accuracies of 96.58\% and 94.95\% on Housing A and B, respectively, followed by ResNet152V2 at 96.49\% and 95.35\%, and EfficientNetB0 at 94.16\% and 91.65\%, respectively. The models' high accuracies across different housings demonstrate a robust solution suitable for cost-effective condition monitoring deployable near sensing platforms, contributing to applied machine learning for edge intelligence and showcasing effective signal processing strategies for handling complex, potentially large-scale vibration data.
\end{abstract}


\section{Introduction}
Rotating machinery such as motors, turbines, gearboxes, and pumps are critical in industry, and their failures can cause catastrophic damage or costly downtime \cite{pr8101217}. Bearing elements are critical components in a wide array of rotating machinery, encompassing applications such as aviation engines\cite{8740504}, high-speed trains\cite{9656160}, helicopters, wind turbines\cite{9028804}, machine tools, and industrial robots. Notably, these bearings are a significant source of failure in such equipment, accounting for more than forty percent of overall malfunctions\cite{9016119}. Condition monitoring and early fault diagnosis are therefore essential for safety and reliability. Various approaches have been proposed for the diagnosis and prognosis of bearings, including model-based methods, data-driven techniques, signal processing-based strategies, knowledge-based systems, active fault diagnosis, and hybrid methodologies\cite{7069265}. Each approach offers distinct advantages and disadvantages depending on the nature of the application and the availability of system information. 

Traditional fault detection methodologies rely on signal processing techniques to extract defect-sensitive features from vibration signals. The Fast Fourier Transform (FFT) and Short-Time Fourier Transform (STFT) are widely used for stationary signal analysis but struggle with non-stationary, noisy industrial larger datasets\cite{10151088}. Wavelet Transform (WT) and Empirical Mode Decomposition (EMD) address non-stationarity by localizing transient features in time-frequency domains, yet they require manual parameter tuning and lack robustness under variable noise conditions\cite{8404129}. While these methods have been successfully used for bearing fault diagnosis, they have notable limitations. Furthermore, conventional techniques can be sensitive to noise and to variations in operating conditions, for example, slight changes in load or speed can mask fault-related frequencies, challenging the robustness of methods like FFT or WT. In many cases, traditional signal processing struggles to accurately identify incipient faults, especially when signals are non-stationary or contaminated by background noise\cite{10931879}.

\begin{figure*}[htbp]  
    \centering
    \includegraphics[width=0.95\textwidth]{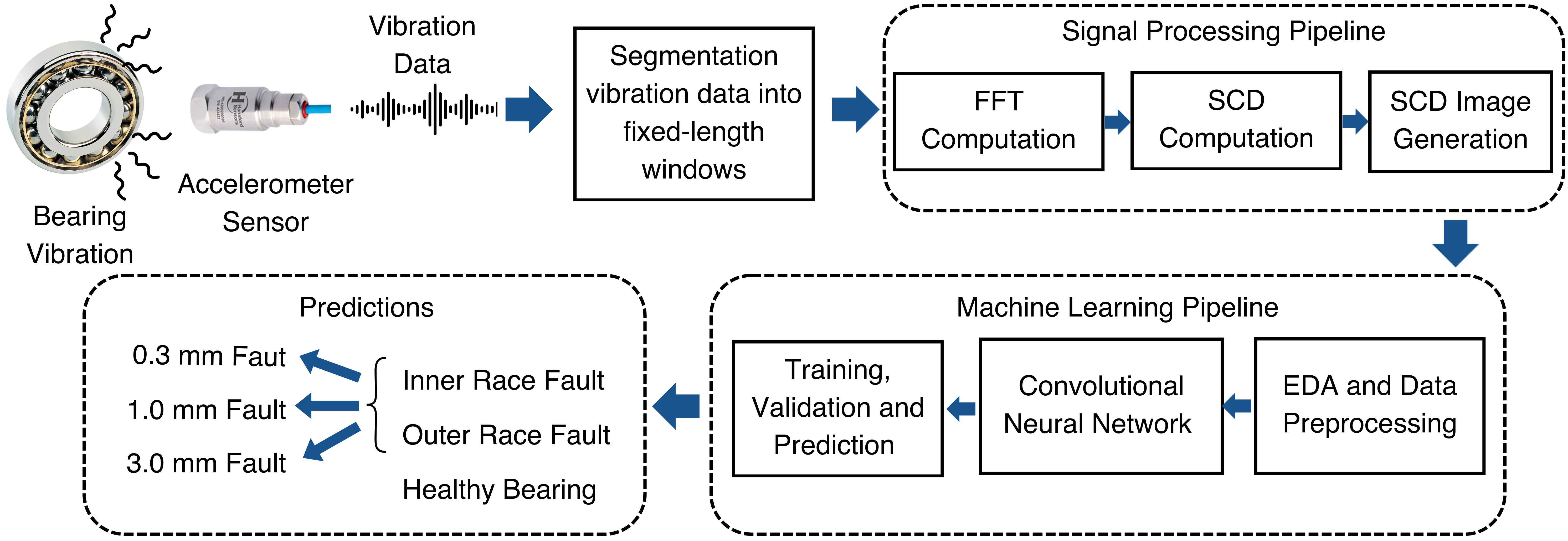}
    \caption{The complete system architecture of the Rotating Machinery Fault Monitoring System}
    \label{sysarchi}
\end{figure*}

In recent years, data-driven deep learning (DL) approaches have achieved remarkable success in fault diagnosis, by automatically learning discriminative features from raw sensor data. Convolutional Neural Networks (CNNs) excel at learning spatial patterns from time-frequency representations\cite{8675278}\cite{9729385}, while recurrent architectures (e.g., LSTMs) model temporal dependencies in vibration sequences\cite{10393124}. Hybrid frameworks, such as CNN-LSTM networks, have achieved state-of-the-art accuracy on benchmark datasets like Case Western Reserve University (CWRU) bearings\cite{8566908}. Recent advancements leverage cyclostationary analysis, to exploit periodic modulation patterns in vibration signals caused by faults. Speed curve estimation method combined with time-angle cyclostationary analysis for bearing diagnosis is proposed which uses the instantaneous frequency of the vibration signal\cite{9831639}. 

However, DL models predominantly use raw vibration signals or basic time-frequency transforms (e.g., STFT, WT) as inputs, neglecting the rich diagnostic information in SCD representations. Furthermore, current DL frameworks degrade significantly under real-world noise, as shown by \cite{JANSSENS2016331} in their study on industrial vibration data. To bridge these gaps, this work proposes integrating SCD imaging with a lightweight deep learning model suitable for deployment on edge intelligence platforms. By fusing cyclostationarity strengths with DL’s adaptive learning, we have enhanced capabilities of machine learning tailored for sensing applications. The key contributions of this paper are:

\begin{itemize}
    \item Novel Hybrid Framework: A fault detection system combining SCD imaging for cyclostationary feature extraction with a computationally efficient CNN architecture. By leveraging cyclostationary signal properties, the proposed SCD+DL pipeline achieves higher fault detection accuracy compared to conventional FFT-based or time-domain methods
    \item Noise-Robust Design: Enhanced generalization under varying signal-to-noise ratios (SNRs) through SCD’s inherent noise suppression properties [16].
    \item Benchmark Dataset Evaluation: The approach is evaluated on a publicly available Rotating Machinery Vibration dataset\cite{JUNG2023109049} that includes normal and faulty bearing conditions under various loads, demonstrating superior accuracy.
\end{itemize}

The remainder of the paper is organized as follows: Section II describes methodology; Section III, signal processing and machine learning; Section IV, experimental results, and finally in Section V, conclusions are drawn with the future research directions.

\section{Methodology}

Fig. \ref{sysarchi} illustrates the complete system architecture of the Rotating Machinery Fault Monitoring System. We have utilized the dataset in \cite{JUNG2023109049} which consists of vibration, acoustic, temperature, and motor current datasets of rotating machines under different load conditions of 0 Nm, 2 Nm and 4 Nm. For our analysis, we have focused on vibration data, which were measured using four accelerometers (\textit{PCB352C34}) placed at two different bearing housings A and B (two adjacent bearing housings) in both x and y directions based on ISO \(10816-1:1995\). The data has been collected at a sampling frequency of 25.6 kHz by keeping the machinery 120 seconds and 60 seconds in normal state, and  faulty state respectively\cite{JUNG2023109049}. The vibration data are stored in binary MATLAB (MAT) and the unit of the data is ‘gravitational constant (g)’ ($1g = 9.80665 ms^{-2}$ ). 

\begin{figure}[tbp]  
    \centering
    \includegraphics[width=0.35\textwidth]{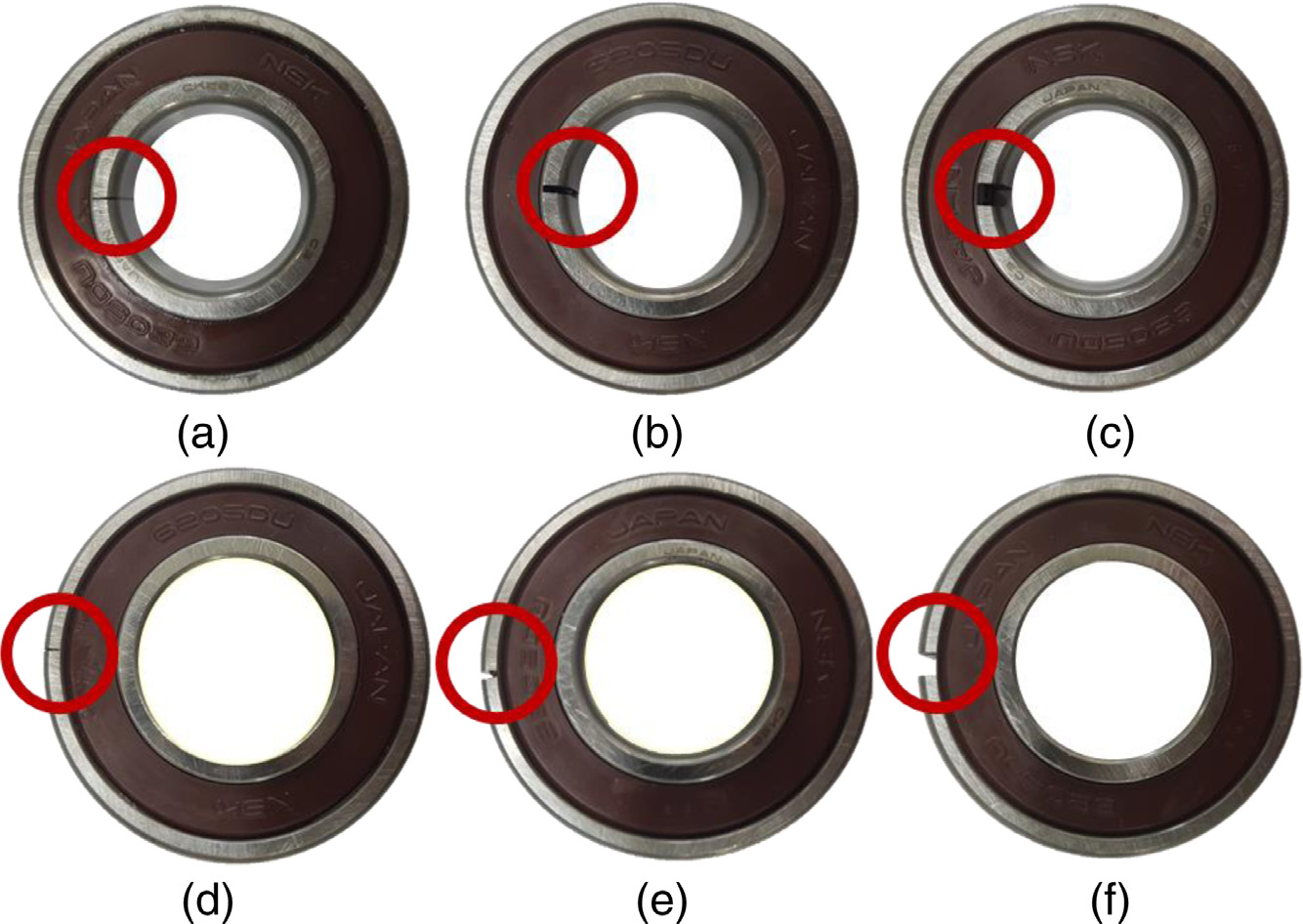}
    \caption{Bearing by crack size: (a) inner race 0.3 mm (small), (b) inner race 1.0 mm (medium), (c) inner race 3.0 mm (large), (d) outer race 0.3 mm (small), (e) outer race 1.0 mm (medium), and (f) outer race 3.0 mm (large) \cite{JUNG2023109049}}
    \label{bearing_crack}
\end{figure}

Initially, the data is segmented into fixed-length windows of 10,000 data samples. Then a signal processing pipeline was introduced to extract the special features of the vibration data obtained by vibration sensors. Signals were analyzed in the frequency domain by utilizing the Fourier transform, particularly the Fast Fourier Transform (FFT). Then the cyclostationary property analysis is performed by spectral correlation density (SCD) calculation followed by SCD image generation. These generated SCD images are used to feed as input data to the deep learning based models to predict the condition of the motor. Mainly, the model predicts whether there's a fault in the bearings or a healthy bearing. Then if it is a faulty bearing, the model is capable of identifying it as an inner race fault or an outer race fault. Further classifying these faults, it can predict the crack size of the bearing into 3 distinct classes as small, medium, and large cracks consisting with crack sizes of  \(0.3 mm\), \(1.0 mm\) and \(3.0 mm\) respectively, for each inner and outer race faults (shown in Fig. \ref{bearing_crack}), making a sum of 6 different classes for faulty conditions. This enables the user to get an idea of the bearing condition to estimate its severity, which is crucial for predictive maintenance and failure prognosis.

\section{Signal Processing and Machine Learning}

In our study, we employ cyclostationary property analysis to enhance signal separability, improving detection accuracy. As the initial step to create a dataset to build the model, these four types of data were taken into separate NumPy arrays. The \textit{SciPy} library from Python was used to load and inspect the data in the original MAT files. For data processing, raw data in MAT files were converted to numpy arrays. To create a dataset to build the classification model, array slicing was used. Here, the numpy array was sliced, resulting in a block of 10,000 elements, and saved as MAT files for further processing. Also, the data file consisted of information for the timestamp as the start value, increment, and the number of values. By using Eq~\ref{eq1}, we calculated a time vector to analyze the signals in the time domain. 

\begin{equation}
    t_i = s + i \cdot \Delta t \quad \text{for } i = 0, 1, 2, \ldots, N-1.
    \label{eq1}
\end{equation}

where, $t_i$, $s$, $\Delta$ and $i$ are, time vector for the signal, start value, increment and number of values in the data file respectively. 

As an example, the \textit{$0Nm\_Normal.mat$} mat file consisted with matrix size of $(7680000,4)$ for vibrational data in the x and y directions of two housings (A and B) of healthy bearing under the 0 Nm load condition. Then four numpy arrays were created for resulting an array size of $(7680000,)$ for each. From these numpy arrays, 10,000 element blocks were sliced and saved as separate MAT files resulting 768 mat files ($7,680,000 / 10,000 = 768$) in total.
To calculate the time vector, for the same mat file \textit{$0Nm\_Normal.mat$} start value, increment and the number of values were given as \(2.6206 \times 10^{-5}\), \(3.906 \times 10^{-5}\) and \(7,680,000\) respectively.

\subsection{Cyclostationarity Property}

Mechanical vibration signals from rotating machinery often exhibit cyclostationary properties due to their inherent periodicity caused by rotating components\cite{9831639}. Unlike stationary signals whose statistical properties remain constant over time, cyclostationary signals possess periodically time-varying statistics, particularly in their mean and autocorrelation functions\cite{ANTONI2007597}.

A key tool for analyzing second-order cyclostationarity is the Cyclic Spectral Correlation Function (CSCF), which reveals how the spectral components of a signal are correlated at specific cyclic frequencies. The CSCF, also known as the Spectral Correlation Density (SCD), is defined as in Eq (\ref{eq:fft5}),

\begin{equation} 
S_x^{\alpha}(f) = \lim_{T \to \infty} \frac{1}{T} , \mathbb{E} \left[ X_T\left(f + \frac{\alpha}{2} \right) \cdot X_T^*\left(f - \frac{\alpha}{2} \right) \right] 
\label{eq:fft5}
\end{equation}

Here, $S_x^{\alpha}(f)$ is the cyclic spectral correlation function at cyclic frequency $\alpha$ and spectral frequency $f$, $X_T(f)$ is the Fourier transform of the signal truncated to a time interval $T$, and $X_T^*(f)$ denotes the complex conjugate of $X_T(f)$. The operator $\mathbb{E}[\cdot]$ represents the statistical expectation, and $\alpha$ is the cyclic frequency, reflecting the periodicity of the signal’s second-order statistics.

\subsubsection{FFT Accumulation Method (FAM)}

The FFT Accumulation Method (FAM) is a computationally efficient technique for estimating the CSCF by averaging frequency-shifted products of short-time Fourier transforms (STFTs) of the input signal\cite{spooner1}. To follow the FAM method, initially the signal was segmented  $x[n]$ into overlapping time windows of length $N'$, using a window function $w[n]$. Then short-time Fourier transform (STFT) was computed for each windowed segment as in Eq (\ref{eq:fft6}),

\begin{equation} 
X_m(f) = \sum_{n=0}^{N'-1} x[n + mL] \cdot w[n] \cdot e^{-j2\pi f n} 
\label{eq:fft6}
\end{equation}

where, $m$ is the segment index, $L$ is the hop size (segment offset), $f$ is the spectral frequency, $X_m(f)$ is the local Fourier transform of the $m$-th segment. Finally, the cyclic correlation for each pair of frequency components separated by the cyclic frequency was computed where $\alpha$ is defined as in Eq (\ref{eq:fft7}),

\begin{equation}
\hat{S}_x^{\alpha}(f) = \frac{1}{M} \sum_{m=0}^{M-1} X_m\left(f + \frac{\alpha}{2}\right) \cdot X_m^*\left(f - \frac{\alpha}{2} \right)
\label{eq:fft7}
\end{equation}
Here $M$ is the total number of segments.

This estimation technique builds a two-dimensional map of cyclic frequency $\alpha$ versus spectral frequency $f$, revealing modulations caused by mechanical faults that are otherwise difficult to detect using traditional FFT-based approaches. The FAM provides a practical balance between resolution and computational efficiency for cyclostationarity property analysis.

\subsection{Deep Learning}

The methodology encompasses data loading, preprocessing, an optimized data pipeline, model training with a custom CNN, and rigorous evaluation. The pipeline is designed to classify SCD images into seven distinct classes, representing various fault conditions and a normal state, as explained in TABLE \ref{tab:class}.

\begin{table}[htbp]
\centering
\caption{Predicting Classes of CNN models}
\label{tab:class}
\begin{tabular}{|l|l|}
\hline
\multicolumn{1}{|c|}{\textbf{Class Name}} & \multicolumn{1}{c|}{\textbf{Information}} \\ \hline
BPFI\_03 & 0.3 mm inner race fault \\ \hline
BPFI\_10 & 1.0 mm inner race fault \\ \hline
BPFI\_30 & 3.0 mm inner race fault \\ \hline
BPFO\_03 & 0.3 mm outer race fault \\ \hline
BPFO\_10 & 1.0 mm outer race fault \\ \hline
BPFO\_30 & 3.0 mm outer race fault \\ \hline
Healthy  & Healthy bearing         \\ \hline
\end{tabular}
\end{table}

This classification is common for both housing A and B used in the analysis. We generated 8,603 SCD images for each of these housing types (A and B). and utilized for the classification moodel. The data was consisting with all the three types of loading conditions mentioned in \cite{JUNG2023109049} as $0\,\text{Nm}$ (No load condition), $2\,\text{Nm}$ and $4\,\text{Nm}$.

The dataset is processed using TensorFlow's image dataset from directory utility, which automates class label inference from the directory structure and resizes all images to a uniform resolution of 224$\times$224 pixels. This standardization ensures compatibility with the CNN input requirements. Labels are one-hot encoded to support multi-class classification, streamlining the subsequent training process.

The SCD images, originally with pixel values in the range [0, 255], are normalized to the range [0, 1] by dividing each pixel value by 255. This accelerates gradient convergence and mitigates numerical instability during training. The preprocessing is seamlessly integrated into the data pipeline using TensorFlow’s \textit{map} function, which applies the scaling transformation as the batches are loaded. This approach optimizes memory usage by avoiding the need to preload the entire dataset, making it scalable for large-scale applications.

The data pipeline leverages TensorFlow’s \textit{tf.data.Dataset} API to ensure efficient and scalable data handling. Key components of the pipeline can be identified as below,

\begin{itemize}
    \item Batching: The dataset is segmented into batches of 32 images, balancing computational efficiency and GPU memory utilization during training.
    \item Shuffling: Random shuffling of the dataset is applied to eliminate biases arising from the order of data presentation, enhancing the model’s generalization capability.
    \item Prefetching: The pipeline employs prefetching to overlap data preparation with model training, reducing GPU idle time and improving throughput.
\end{itemize}

To ensure the model is robust and avoids overfitting, the dataset is divided into three distinct subsets: training, validation, and test sets in the ratios of seventy percent (\(70\%\)), twenty percent (\(20\%\)), and ten percent (\(10\%\)) respectively. Training is conducted in a GPU-accelerated environment to leverage parallel processing capabilities, significantly reducing computation time. The model is trained over 15 epochs , with an early stopping mechanism implemented to terminate training if the validation loss ceases to improve over a predefined number of epochs. This strategy mitigates overfitting and conserves computational resources. The models used for the system are explained below.

\subsubsection{Custom CNN}
This CNN consists of three convolutional layers with increasing filter counts (16, 32, and 64), each followed by a max-pooling layer to reduce spatial dimensions while preserving critical features. The convolutional layers employ 3x3 kernels with 'same' padding and ReLU activation functions to introduce non-linearity and enhance feature learning. After the convolutional and pooling layers, the feature maps are flattened and passed through a dense layer with 128 units and ReLU activation, followed by a dropout layer with a rate of 0.5 to mitigate overfitting. The final layer is a softmax activation with seven units corresponding to the bearing conditions. This custom design allows the model to effectively capture the unique characteristics of SCD images while maintaining computational efficiency.

\subsubsection{EfficientNetB0}
EfficientNetB0 was used for classification by leveraging pre-trained weights from ImageNet and replacing its top layers with a custom classification head tailored to the seven-class problem. The base model was frozen (set to non-trainable) to retain its general feature extraction capabilities while focusing training on the new head, which consisted of global average pooling to reduce spatial dimensions, batch normalization for improved training stability, dropout with a rate of 0.2 for regularization, and a dense layer with softmax activation for the seven bearing conditions. This approach allowed EfficientNetB0 to efficiently process the 2D SCD images while minimizing computational overhead.

\subsubsection{ResNet152V2}
ResNet152V2 is a deep residual network with 152 layers, designed to address the challenges of training very deep networks through the use of residual connections. In this research, ResNet152V2 was employed with pre-trained ImageNet weights. The base model was frozen, and a custom classification head was added, comprising global average pooling to aggregate spatial information from the feature maps, a dense layer with 512 units and ReLU activation for further feature processing, dropout with a rate of 0.5 to prevent overfitting, and a final dense layer with softmax activation for the seven bearing conditions. The depth of ResNet152V2 enables it to learn hierarchical features effectively, making it well-suited for complex tasks like fault diagnosis, where intricate patterns may be present in the data.

\section{Experimental Results}

\begin{figure*}[tbp]
\centering 
\includegraphics[width=0.8\textwidth, keepaspectratio]{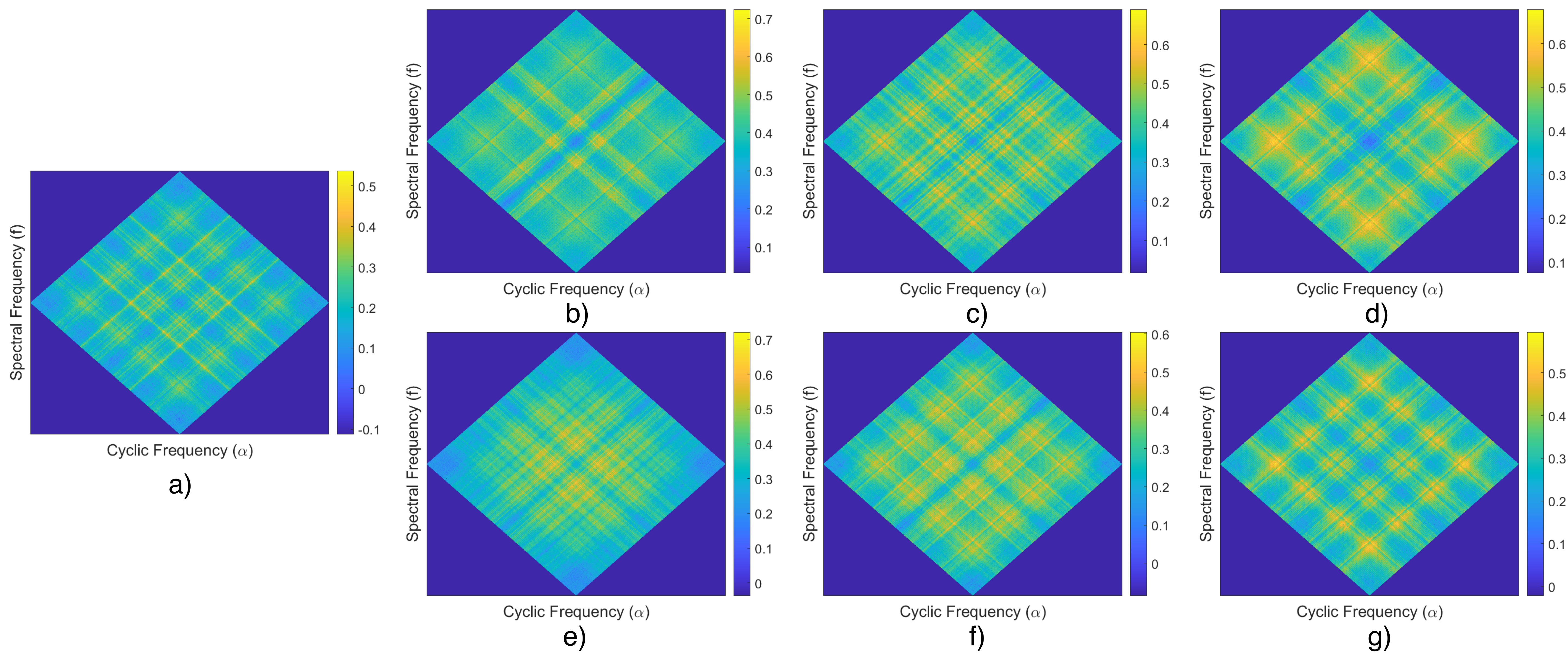} 
\caption{2D SCD patterns obtained for different types of vibration signals for bearings a)Healthy Bearing b)0.3 mm inner race fault c)1 mm inner race fault d)3 mm inner race fault e)0.3 mm outer race fault f)1 mm outer race fault g)3 mm outer race fault} 
\label{dsp2} 
\end{figure*}


\subsection{Cyclostationarity Property Analysis}

Traditional FFT analysis offers insights into the power spectral density (PSD). However, for bearing fault diagnosis, FFT has significant limitations, particularly with non-stationary signals. Bearing vibration signals are inherently cyclostationary due to the periodic nature of rotating elements
and their interaction with faults. Cyclostationary analysis, particularly through SCD, addresses this by capturing the periodic modulation through the correlation between different frequency components over time. Unlike FFT, which only provides the overall frequency content, SCD reveals hidden periodicities by identifying cycle frequencies (\(\alpha\)) where the signal exhibits cyclostationarity. Fig. \ref{dsp2} shows SCD patterns for different bearing conditions as 2D plots. 


In 2D SCD plots (Fig. \ref{dsp2}), the magnitude of the spectral correlation(\(S_x\)) is represented as a color map over the \(f-\alpha\) plane, with color intensity indicating the strength of correlation (e.g., dark blue for low, yellow for high). Here, in Fig. \ref{dsp2} a), healthy bearing illustrates uniform, symmetrical patterns with moderate intensity, lacking prominent peaks outside the main diagonal, indicating no significant cyclic components associated with faults. When analyzing faulty conditions, referred to in Fig. \ref{dsp2} b), c), and d), increased intensity (yellow to orange) and more complex grid patterns with pronounced diagonal lines suggest periodic impulses due to faults. As fault severity increases, the pattern becomes denser, with higher correlation at specific cyclic frequencies.

\subsection{Model Performance Analysis}

We built three CNN models, \textit{EfficientNetB0}, \textit{ResNet152V2}, and a custom CNN, to classify bearing conditions into seven classes (refer to TABLE \ref{tab:class}) and the complete performance comparison of these models related to housing A and B is illustrated in TABLE \ref{main-table}. The models were evaluated on precision, recall, F1-score, and overall accuracy performance metrics.







The custom CNN achieves the highest overall accuracy at \(96.58\%\) and \(95.35\%\) for housing A and B, respectively. This suggests that the custom CNN, likely tailored to the specific characteristics of the SCD images, outperforms pre-trained models, which may not be optimized for this domain. \textit{ResNet152V2}, with its deeper architecture and residual connections, performs closely, indicating that depth and feature reuse are beneficial for capturing complex fault signatures. \textit{EfficientNetB0}, designed for efficiency, lags behind, possibly due to its lighter architecture being less suited for the nuanced patterns in SCD images.

All models exhibit near-perfect performance for the healthy class (F1-scores \(~0.98-0.99\)), indicating that the SCD images for healthy bearings are distinctly separable from faulty states, likely due to smoother, less modulated spectra. Smaller faults (0.3 mm, 1.0 mm), for BPFI\_03, BPFI\_10, BPFO\_03, and BPFO\_10, F1-scores are generally above 0.95 across all models, with perfect precision in several cases (e.g., \textit{ResNet152V2} for BPFI\_03, BPFI\_10). This suggests that smaller faults produce clear, distinguishable signatures in the SCD images, likely due to localized impacts at characteristic frequencies.

For larger faults (3.0 mm), performance decreases for BPFI\_30 and BPFO\_30, particularly in precision. For BPFI\_30, \textit{EfficientNetB0} and \textit{ResNet152V2} show lower precision (0.80 and 0.78, respectively), with F1-scores of 0.88 and 0.87, indicating false positives (samples from other classes misclassified as BPFI\_30). For BPFO\_30, the Custom CNN has lower precision (0.82) and F1-score (0.90), suggesting similar issues. This trend aligns with the hypothesis that larger faults generate more complex, broadband vibration patterns, potentially overlapping with other fault types or the healthy state, complicating classification.

The recall for BPFI\_30 is high, indicating that most actual BPFI\_30 samples are correctly identified, but the low precision suggests misclassifications from other classes. Similarly, for BPFO\_30, the Custom CNN has higher recall (0.99), but lower precision (0.82), pointing to false positives.

\begin{table*}[htbp]
\centering
\caption{Performance comparison of EfficientNetB0, ResNet152V2, and Custom CNN on Housing A and Housing B datasets across various fault types}
\label{main-table}
\begin{tabular}{@{}llcccccccccccccc@{}}
\toprule
                                         &                    & \multicolumn{7}{c}{Housing A}                  & \multicolumn{7}{c}{Housing B}                  \\ 
 &
   &
  \multicolumn{3}{c}{\textbf{BPFI (mm)}} &
  \multicolumn{3}{c}{\textbf{BPFO (mm)}} &
  \multirow{2}{*}{\textbf{Normal}} &
  \multicolumn{3}{c}{\textbf{BPFI (mm)}} &
  \multicolumn{3}{c}{\textbf{BPFO (mm)}} &
  \multirow{2}{*}{\textbf{Normal}} \\
CNN Model                                & Metrics            & 0.3  & 1.0  & 3.0  & 0.3  & 1.0  & 3.0  &      & 0.3  & 1.0  & 3.0  & 0.3  & 1.0  & 3.0  &      \\ \midrule
\multirow{4}{*}{\textbf{EfficientNetB0}} & \textbf{Precision} & 0.98 & 0.98 & 0.80 & 1.00 & 0.95 & 0.96 & 0.99 & 0.98 & 0.96 & 1.00 & 0.90 & 0.98 & 0.91 & 0.98 \\
                                         & \textbf{Recall}    & 0.94 & 0.95 & 0.97 & 0.96 & 0.90 & 0.92 & 0.99 & 0.93 & 0.96 & 0.94 & 0.98 & 0.93 & 0.96 & 0.98 \\
                                         & \textbf{F-1 Score} & 0.96 & 0.97 & 0.88 & 0.98 & 0.93 & 0.94 & 0.99 & 0.96 & 0.96 & 0.97 & 0.94 & 0.95 & 0.93 & 0.98 \\
                                         & \textbf{Acuracy}   & \multicolumn{7}{c}{\textbf{94.16\%}}           & \multicolumn{7}{c}{\textbf{91.65\%}}           \\
\multirow{4}{*}{\textbf{ResNet152V2}}    & \textbf{Precision} & 1.00 & 1.00 & 0.78 & 0.95 & 1.00 & 0.98 & 1.00 & 0.98 & 1.00 & 1.00 & 0.99 & 0.97 & 0.83 & 1.00 \\
                                         & \textbf{Recall}    & 0.93 & 0.95 & 1.00 & 0.94 & 0.92 & 0.98 & 0.97 & 0.94 & 0.95 & 0.95 & 0.97 & 0.96 & 1.00 & 0.99 \\
                                         & \textbf{F-1 Score} & 0.96 & 0.97 & 0.87 & 0.94 & 0.96 & 0.98 & 0.99 & 0.96 & 0.98 & 0.97 & 0.98 & 0.97 & 0.91 & 0.99 \\
                                         & \textbf{Acuracy}   & \multicolumn{7}{c}{\textbf{96.49\%}}           & \multicolumn{7}{c}{\textbf{94.95\%}}           \\
\multirow{4}{*}{\textbf{Custom CNN}}     & \textbf{Precision} & 1.00 & 1.00 & 0.98 & 0.99 & 1.00 & 0.82 & 0.97 & 0.95 & 1.00 & 1.00 & 0.84 & 0.98 & 1.00 & 0.97 \\
                                         & \textbf{Recall}    & 0.94 & 0.96 & 0.94 & 0.95 & 0.95 & 0.99 & 0.99 & 0.93 & 0.94 & 0.95 & 0.97 & 0.93 & 0.94 & 0.99 \\
                                         & \textbf{F-1 Score} & 0.97 & 0.98 & 0.96 & 0.97 & 0.97 & 0.90 & 0.98 & 0.94 & 0.97 & 0.97 & 0.9  & 0.96 & 0.97 & 0.98 \\
                                         & \textbf{Acuracy}   & \multicolumn{7}{c}{\textbf{96.58\%}}           & \multicolumn{7}{c}{\textbf{95.35\%}}           \\ \bottomrule 
\end{tabular}
\end{table*}

Furthermore, a clear trend is observed for all three models consistently achieving higher accuracies when classifying data from Housing A compared to Housing B, with differences ranging from approximately \(1.5\%\) to \(2.5\%\) across the models. The models exhibit peak performance on Housing A, with the custom CNN achieving the highest accuracy at \(96.58\%\), followed closely by \textit{ResNet152V2} at \(96.49\%\), and \textit{EfficientNetB0} at \(94.16\%\).The primary reason for the difference in performance between Housing A and Housing B lies in the experimental setup and the nature of the vibration data collected from each housing.

However, the models perform exceptionally well regardless of whether the vibration data comes from Housing A or B. For instance, accuracies like \(94.95\%\) (Custom CNN) and \(95.35\%\) (ResNet152V2) on Housing B, despite a slight drop compared to Housing A, indicate that the models can effectively detect faults even when sensors are not placed directly on the faulty part. This robustness makes them highly suitable for detecting vibration-based faults at any location in rotating machinery, especially in real-world scenarios where optimal sensor placement may not always be feasible due to accessibility, cost, or safety constraints.

The models strong performance on both Housing A and Housing B provides confidence that they can detect faults throughout the rotating machinery, not just at the exact fault location. Even with indirect data from Housing B achieving over \(91\%\) accuracy, the models can infer fault conditions from secondary vibration effects. This capability is invaluable for comprehensive condition monitoring, ensuring the health of the entire system is assessed effectively.

\section{Conclusion}

The study's findings underscore the effectiveness of deep learning for bearing fault diagnosis, with SCD images proving superior to FFT by capturing cyclostationary properties, enhancing diagnostic precision. The models' high accuracies on both housings highlight their versatility, enabling cost-effective condition monitoring adaptable to diverse sensing platform configurations. The Custom CNN's efficiency and top performance suggest its strong potential for real-time edge intelligence applications requiring rapid diagnostics, while \textit{ResNet152V2} and \textit{EfficientNetB0} offer depth and balance, respectively, for broader scenarios. These results validate deep learning's role in predictive maintenance and automated fault diagnosis, with potential applications in wind turbines, gearboxes, and manufacturing, where robust fault detection is paramount. Future research could explore multi-sensor fusion, fine-tuning pre-trained models, and extending to varying speed conditions to further enhance applicability, ensuring comprehensive fault diagnosis in diverse industrial contexts.


\bibliographystyle{ieeetr}

\bibliography{references}

\end{document}